\documentclass[fp,twocolumn]{jpsj3}
\usepackage{txfonts}
\usepackage{color}
\usepackage{setspace}  

\title{Cu-spin Correlation in the Electron-overdoped High-$T_{\rm c}$ Cuprate Thin Films of La$_{2-x}$Ce$_x$CuO$_4$ Probed by Low-energy Muons}

\author{SangEun Park,$^1$ Yuma Kawai,$^1$ Andreas Suter,$^2$ Hirotaka Okabe,$^3$ Jumpei G. Nakamura,$^3$ Hideki Kuwahara,$^1$ Zaher Salman,$^2$ Thomas Prokscha,$^2$ Ryosuke Kadono,$^3$ and Tadashi Adachi$^1$\thanks{t-adachi@sophia.ac.jp}}
\inst{$^1$Department of Engineering and Applied Sciences, Sophia University, 7-1 Kioi-cho, Chiyoda-ku, Tokyo 102-8554, Japan \\
$^2$PSI Center for Neutron and Muon Sciences, 5232 Villigen PSI, Switzerland \\
$^3$Institute of Materials Structure Science, High Energy Accelerator Research Organization (KEK-IMSS), 1-1 Oho, Tsukuba 305-0801, Japan} 

\abst{We investigated the Cu-spin correlation in the overdoped regime of the electron-doped high-$T_{\rm c}$ cuprate thin films of La$_{2-x}$Ce$_x$CuO$_4$, changing the reduction condition from muon spin relaxation using low-energy muons. 
The Cu-spin correlation developed at low temperatures for optimally reduced films with $x=0.13$ as well as $x=0.17$ where the superconductivity was almost suppressed. 
These results are contrary to those observed in the hole-doped high-$T_{\rm c}$ cuprates where the development of the antiferromagnetic Cu-spin correlation disappears together with the suppression of superconductivity. 
The Cu-spin correlation developed at low temperatures in $x=0.17$ may be understood in terms of antiferromagnetism, but it may be related to a ferromagnetic order recently suggested in the nonsuperconducting heavily overdoped La$_{2-x}$Ce$_x$CuO$_4$ with $x \sim 0.18$.}

\begin{document}
\maketitle

\section{Introduction}

In high-$T_{\rm c}$ cuprate superconductors, the doping of carriers into parent antiferromagnetic (AF) Mott insulators leads to the appearance of superconductivity. 
Previous neutron-scattering~\cite{birgeneau,fujita} and muon-spin-relaxation ($\mu$SR)~\cite{koike-adachi} results suggest that AF spin fluctuations, observed in the underdoped and optimally doped regimes of hole-doped cuprates, are crucial for high-$T_{\rm c}$ superconductivity.
In electron-doped cuprates, it had been believed that the superconductivity appeared through not only electron doping but also the reduction annealing bringing about the removal of excess oxygen in an as-grown sample. 
Formerly, however, it was reported that in the so-called T'-type electron-doped cuprate thin films~\cite{tsukada,matsumoto} and polycrystals,~\cite{asai,takamatsu} through the appropriate removal of excess oxygen, superconductivity appeared in the parent and underdoped samples. 
The results of our previous transport and $\mu$SR studies of the parent and underdoped T'-cuprates suggested that superconductivity appeared under a strong electron correlation.~\cite{adachi-jpsj,adachi-musr} 
To explain the superconductivity in the parent T'-cuprates, mainly two candidates have been proposed: an electronic structure model including the collapse of the charge-transfer gap~\cite{adachi-jpsj,adachi-musr,adachi-condmat} and the excess electron doping by the removal of oxygen.~\cite{horio-natcom,horio-prl,wei,lin}
Moreover, the electron pairing mediated by spin fluctuations has been proposed for the superconductivity in the parent T'-cuprates.~\cite{ohashi,taro,lee}

In the overdoped regime where the superconducting transition temperature $T_{\rm c}$ decreases with carrier doping, neutron-scattering experiments of the hole-doped cuprate La$_{2-x}$Sr$_x$CuO$_4$ revealed that low-energy AF fluctuations disappeared concomitant with the suppression of superconductivity.~\cite{wakimoto}
The results of $\mu$SR studies of the Zn-substituted La$_{2-x}$Sr$_x$Cu$_{1-y}$Zn$_y$O$_4$ in the overdoped regime also suggested an intimate relation between superconductivity and stripe-like AF fluctuations.~\cite{risdiana} 
In the nonsuperconducting heavily overdoped regime, a ferromagnetic order/fluctuation has been proposed theoretically~\cite{kopp,maier} and experimentally~\cite{sonier,kurashima,komiyama,adachi-materials}, suggesting the suppression of superconductivity by ferromagnetic fluctuations.

For the electron-doped cuprates in the overdoped regime, inelastic neutron-scattering experiments of Pr$_{1-x}$LaCe$_x$CuO$_4$ (PLCCO)~\cite{fujita-plcco} revealed that the integrated intensity of the dynamical spin susceptibility $\chi^"$($\omega$) corresponding to low-energy AF fluctuations decreased gradually with overdoping but seemed to be finite in the nonsuperconducting heavily overdoped regime, which is contrary to the results of hole-doped La$_{2-x}$Sr$_x$CuO$_4$.~\cite{wakimoto} 
Moreover, our $\mu$SR measurements of overdoped PLCCO~\cite{malik} revealed that the development of the Cu-spin correlation weakened with overdoping and disappeared around the end point of the superconducting regime, suggesting an intimate relation between the development of the Cu-spin correlation and superconductivity in PLCCO, similarly to the hole-doped cuprates.~\cite{wakimoto} 
However, predominant effects of Pr$^{3+}$ moments mask the behavior of Cu spins on $\mu$SR spectra, and therefore, an investigation using T'-cuprates without rare-earth moments is desired for further understanding.

The electron-doped T'-cuprate La$_{2-x}$Ce$_x$CuO$_4$ (LCCO) without rare-earth moments, obtained only in a thin-film form, exhibits the highest $T_{\rm c}$ of 25 K among T'-cuprates,~\cite{sawa} therefore making it suitable for investigating the Cu-spin correlation and superconductivity. 
Previous $\mu$SR measurements using low-energy muons on LCCO thin films revealed that an AF order was formed for $x \le 0.08$ inside the film and $x \le 0.10$ near the surface of the film.~\cite{saadaoui} 
On the other hand, the angle-dependent magnetoresistance of LCCO suggested that the AF order survived above $x=0.12$.~\cite{jin} 
Therefore, the details of the dependence of the Cu-spin correlation on the electron concentration and its difference between the surface and inside the film in overdoped LCCO have not yet been clarified.

In this study, we investigated the Cu-spin correlation in the overdoped LCCO thin films with $x=0.13$ (nearly optimally doped regime) and 0.17 (close to the end point of the superconducting regime) by $\mu$SR, changing the reduction condition and implantation energy of muons to probe near the surface and inside the film. 
To understand the electronic state, we also performed the resistivity and Hall measurements of LCCO, changing the electron concentration and reduction condition.

\section{Experimental Methods}

LCCO thin films were deposited on (001) SrTiO$_3$ (STO) substrates by pulsed-laser deposition (PLD) utilizing a Nd:YAG laser as an exciting light source. 
In general, high-$T_{\rm c}$ cuprate thin films are prepared by PLD using a KrF excimer laser with a wavelength of 248 or 193 nm. 
In this study, the 3rd harmonic of a Nd:YAG laser with a wavelength of 355 nm was used. 
The conditions of LCCO film growth were as follows: the substrate temperature was $600 - 650$ $^\circ$C, the partial oxygen pressure was $100 - 130$ Pa, and the laser energy density was 83 mJ/mm$^2$. 
The thickness of grown films was estimated, using a confocal laser microscope, to be 1200 nm for $x = 0.13$ and 800 nm for $x=0.17$. 
After growth, post-reduction annealing was performed in the PLD chamber at the same temperature as the deposition temperature under $10^{-4} - 10^{-5}$ Pa for $15 - 35$ min. 
By this annealing, the following samples were prepared: less-reduced (under-reduced) sample, optimally reduced sample where $T_{\rm c}$ is maximum, and excessively reduced (over-reduced) sample. 

The crystal structure and c-axis lattice constant of the film were confirmed by X-ray diffraction analysis.
The electrical and Hall resistivities were measured using a commercially available apparatus (Quantum Design, PPMS).
For these measurements, thin films were prepared to have a surface area of $0.4 \times 3.6$ mm$^2$ with a masking sheet, in accordance with the four- or six-probe method for the electrical and Hall resistivity measurements.

$\mu$SR measurements using low-energy muons were performed at the MuE4 beam line~\cite{prokscha} of the Paul Scherrer Institute in Switzerland. 
In this beam line, the implantation depth of the muon can be controlled from ten to a few hundred nm from the surface by adjusting the energy of muons in fine steps. 
In this study, the implantation depth from the surface of the film was changed from 20 (near the surface) to 110 nm (deep inside the film) by controlling the implantation energy between 3 and 24 keV.
For the $\mu$SR measurements, four films with an area of $10 \times 10$ mm$^2$ were prepared for each composition to cover a total area of $20 \times 20$ mm$^2$, where 90\% of the muons stop.

\section{Results}
\begin{figure}
\begin{center}
\includegraphics[width=1.0\linewidth]{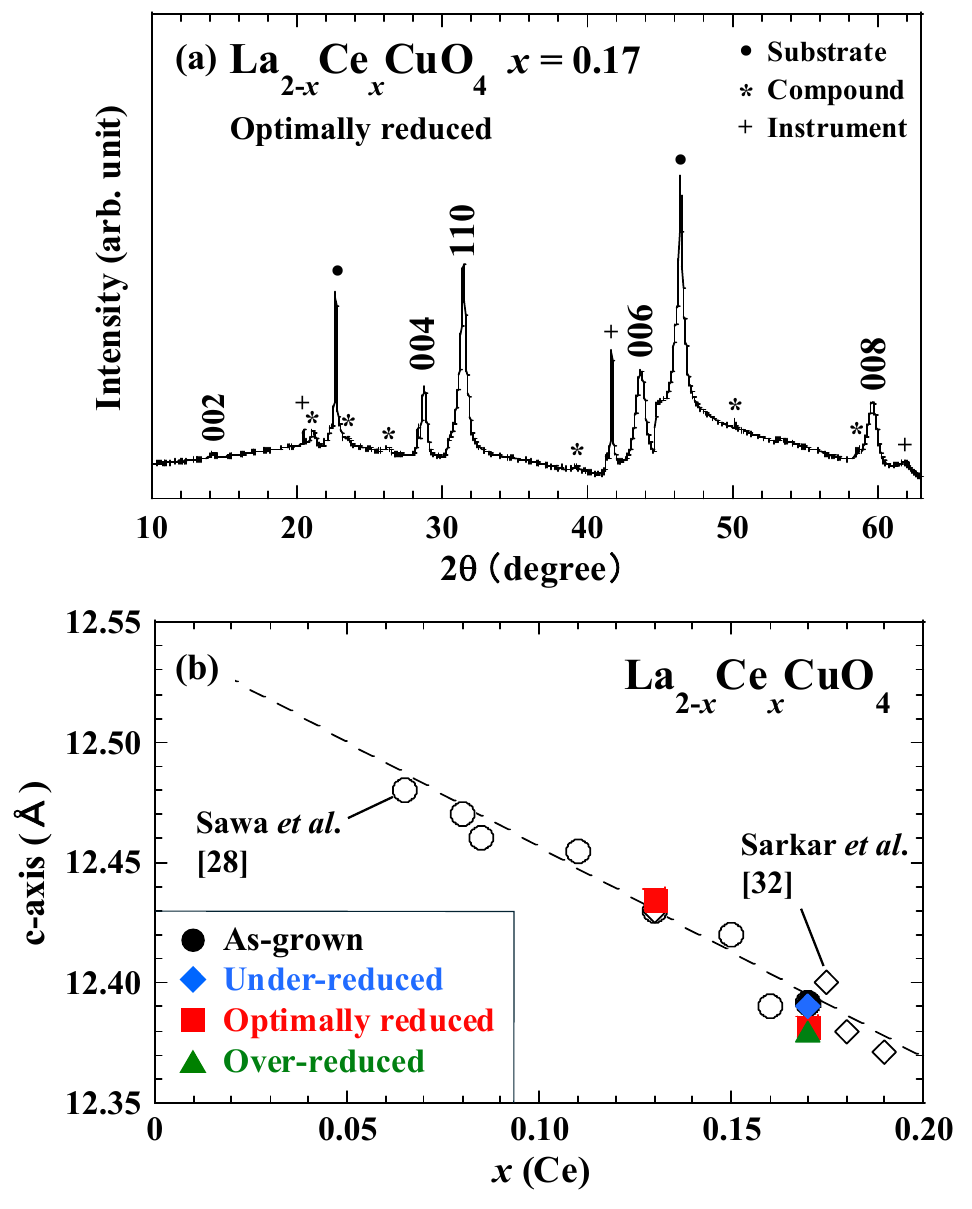}
\caption{(Color online) (a) X-ray diffraction pattern of optimally reduced La$_{2-x}$Ce$_x$CuO$_4$ with $x=0.17$ and (b) Ce concentration dependence of the c-axis lattice constant of La$_{2-x}$Ce$_x$CuO$_4$, together with the previous results. ~\cite{sawa,sarkar}}
\label{f1}
\end{center}
\end{figure}

Figure 1(a) shows the X-ray diffraction pattern of optimally reduced LCCO with $x=0.17$. 
Although the (00l) peaks of LCCO are observed, the (110) peak is also apparent. 
Therefore, the present sample contains domains with both c-axis and (110) orientations, which is probably due to the films having a rather large thickness of $\sim$1000 nm.
The estimated c-axis lattice constants are plotted in Fig. 1(b).
It is found that the present data are almost identical to the previous results.~\cite{sawa,sarkar}
For $x=0.17$, the c-axis length decreases upon reduction annealing, suggesting the removal of excess oxygen~\cite{matsumoto}.
Note that the coexistence of c-axis-oriented and (110)-oriented domains in the thin film may affect the initial asymmetry, i.e., the evaluated magnetic volume fraction in the $\mu$SR results. 
However, it is considered to have no effect on the relaxation rate of muon spins, which reflects the development of a spin correlation.
This is because the muon stopping site is independent of the orientation of the domain.

\begin{figure}
\begin{center}
\includegraphics[width=1.0\linewidth]{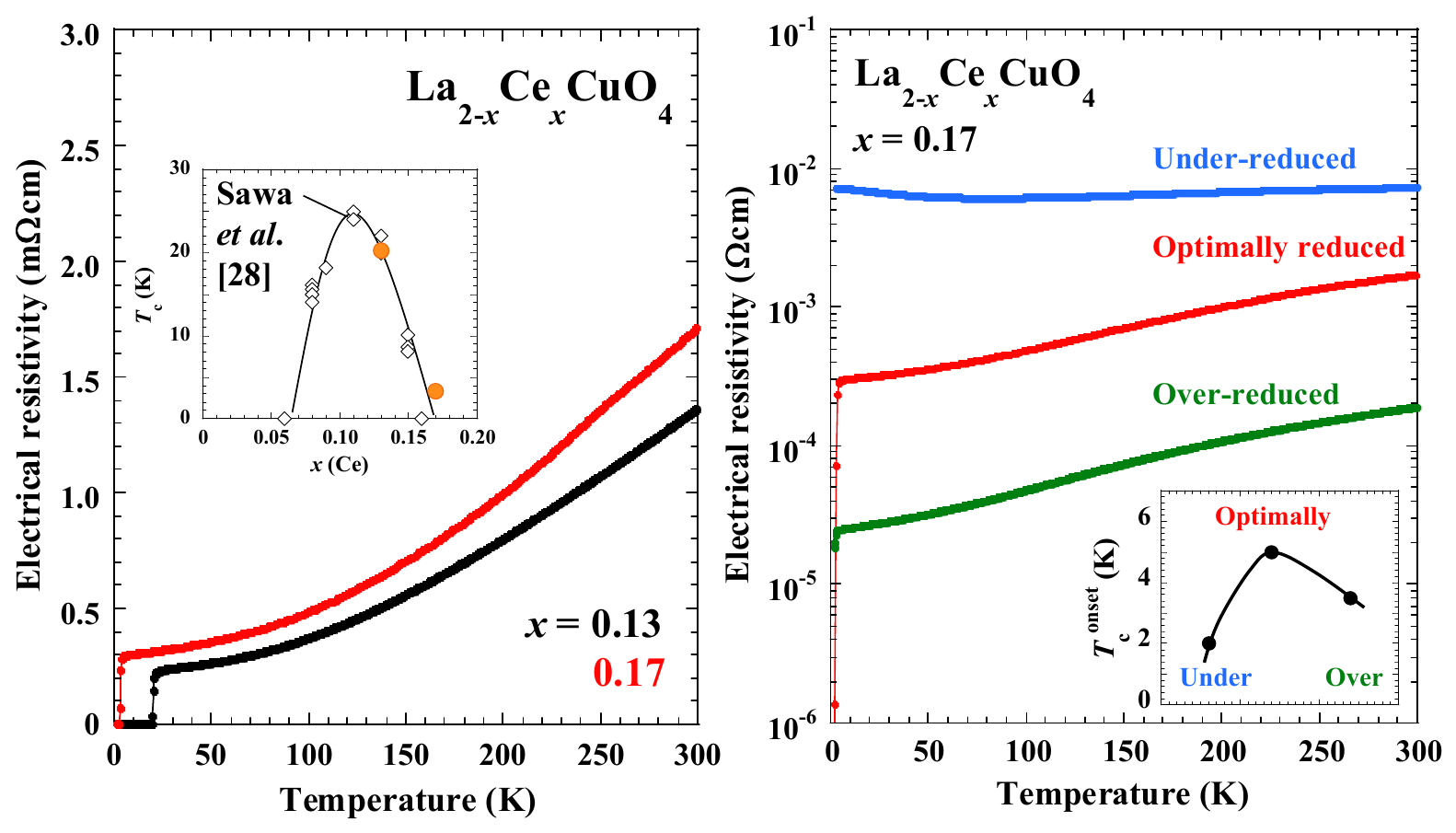}
\caption{(Color online) (a) Temperature dependence of the electrical resistivity of optimally reduced La$_{2-x}$Ce$_x$CuO$_4$ with $x=0.13$ and 0.17. The inset shows the Ce concentration dependence of $T_{\rm c}$, defined as the midpoint temperature of the superconducting transition, together with the previous results.~\cite{sawa} (b) Temperature dependence of the electrical resistivity of La$_{2-x}$Ce$_x$CuO$_4$ with $x=0.17$ under various reduction conditions. The inset shows the change in the onset $T_{\rm c}$, $T_{\rm c}^{\rm onset}$, depending on the reduction condition.}
\label{f2}
\end{center}
\end{figure}

Figure 2(a) shows the temperature dependence of the electrical resistivity of optimally reduced LCCO with $x=0.13$ and 0.17. 
Both samples exhibit metallic behaviors and superconducting transition at low temperatures.
Note that the resistivity at room temperature is higher than those formerly reported,~\cite{sawa} which is probably due to the incomplete epitaxial growth of the present films.
The $T_{\rm c}$'s, defined as the temperature at 50\% of the normal-state resistivity, are plotted in the inset of Fig. 2(a) and are almost identical to those formerly reported.~\cite{sawa}
The resistivity strongly depends on the reduction condition, as shown in Fig. 2(b).
For $x=0.17$, the under-reduced sample exhibits an upturn at low temperatures, while in the optimally reduced sample the resistivity is lower than that of the under-reduced one and exhibits a metallic and superconducting transition, as mentioned above.
The over-reduced sample exhibits the lowest resistivity among the three samples.
However, as shown in the inset of Fig. 2(b), the onset $T_{\rm c}$ is lower than that of the optimally reduced sample.
These results clearly demonstrate that the reduction renders the system conducting owing to the removal of excess oxygen, and excess reduction brings about the destruction of superconductivity owing to the removal of oxygen in the CuO$_2$ plane.

\begin{figure}
\begin{center}
\includegraphics[width=1.0\linewidth]{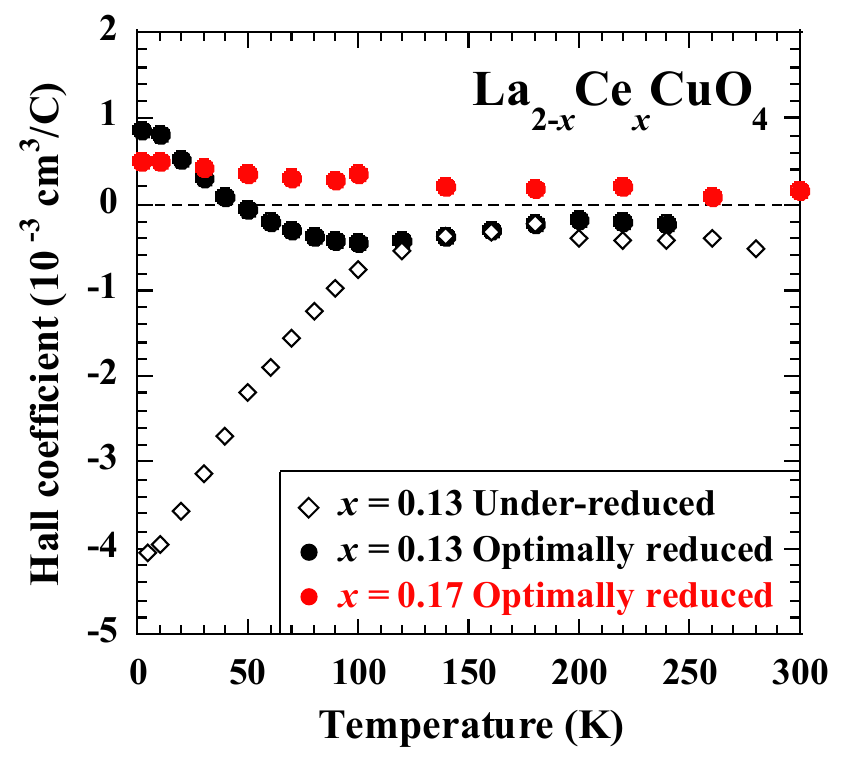}
\caption{(Color online) Temperature dependence of the Hall coefficient of under- and optimally reduced La$_{2-x}$Ce$_x$CuO$_4$ with $x=0.13$ and 0.17.}
\label{f3}
\end{center}
\end{figure}

The temperature dependence of the Hall coefficient $R_{\rm H}$ of LCCO is shown in Fig. 3. 
The $R_{\rm H}$ of under-reduced LCCO with $x=0.13$ is negative below room temperature and decreases monotonically with decreasing temperature at low temperatures. 
On the other hand, the $R_{\rm H}$ of optimally reduced LCCO with $x=0.13$ is almost identical to that of the under-reduced sample at high temperatures, while $R_{\rm H}$ exhibits a sign change at low temperatures.
The $R_{\rm H}$ of optimally reduced LCCO with $x=0.17$ is positive and exhibits weak temperature dependence below room temperature.
These $x$-dependent and reduction-dependent behaviors are consistent with the previous results of LCCO~\cite{sawa} and Pr$_{2-x}$Ce$_x$CuO$_4$.~\cite{dagan,gauthier}

\begin{figure}
\begin{center}
\includegraphics[width=1.0\linewidth]{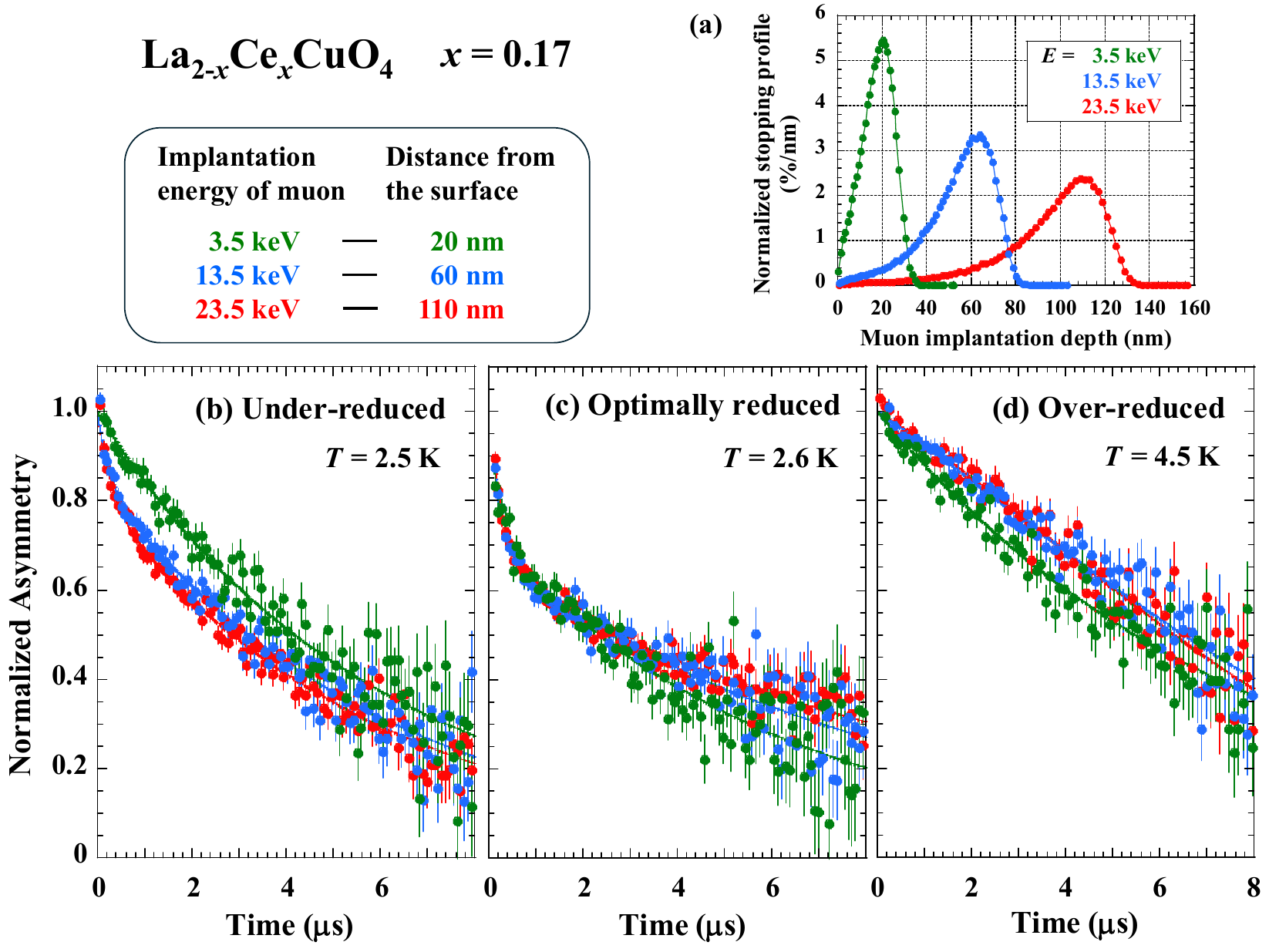}
\caption{(Color online) Zero-field $\mu$SR time spectra of (b) under-, (c) optimally, (d) over-reduced La$_{2-x}$Ce$_x$CuO$_4$ with $x=0.17$, changing the implantation energy of muon. The calculated stopping profile of muons for each implantation energy in the film is shown in (a).}
\label{f4}
\end{center}
\end{figure}

Figure 4 shows zero-field (ZF) $\mu$SR time spectra at the base temperature of La$_{2-x}$Ce$_x$CuO$_4$ ($x=0.17$) with various implantation energies of muons.
As shown in Fig. 4(a), the muon implantation energies of 3.5, 13.5, and 23.5 keV correspond to the maximum muon stopping probabilities of 20 nm (near the surface), 60 nm, and 110 nm (deep inside the film) from the surface of the film, respectively. 
For the optimally reduced sample, while the spectra in a long-time region seem to depend more or less on the implantation energy, the spectra in a short-time region are independent of the implantation energy, indicating that the magnetic state is basically identical throughout the film. 
It is intriguing that the spectrum at 20 nm exhibits a slower (faster) relaxation of muon spins than that at 110 nm for under-reduced (over-reduced) samples. 
The reason for these contrasting results is discussed later. 
For each sample, the spectra at 60 and 110 nm overlap with one another, suggesting that the magnetic state is homogeneous above 60 nm. 
Therefore, the spectra at 23.5 keV (deep inside the film) are focused on the following.

\begin{figure}
\begin{center}
\includegraphics[width=1.0\linewidth]{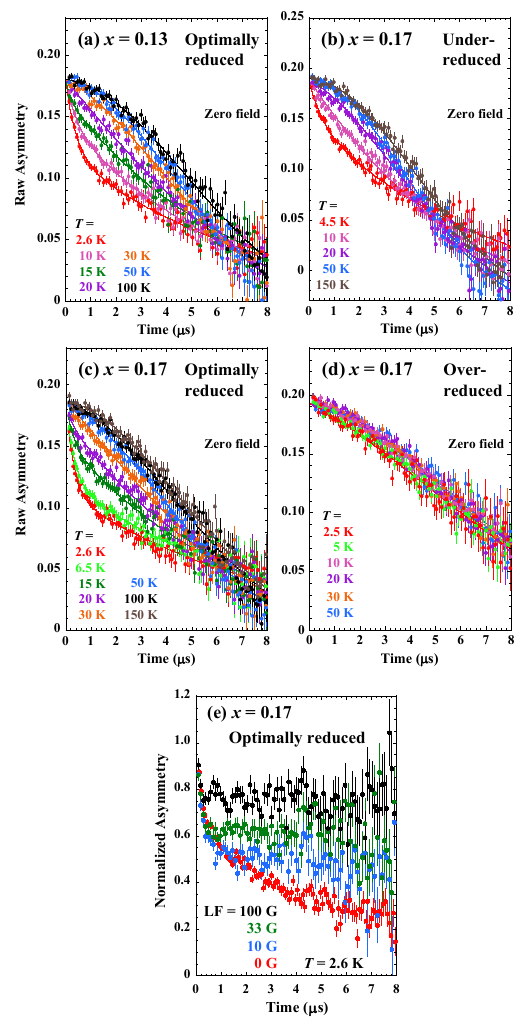}
\caption{(Color online) (a)-(d) Zero-field $\mu$SR time spectra of under-, optimally, and over-reduced La$_{2-x}$Ce$_x$CuO$_4$ with $x=0.13$ and 0.17 obtained at 23.5 keV. (e) Longitudinal-field $\mu$SR time spectra at 2.6 K for optimally reduced La$_{2-x}$Ce$_x$CuO$_4$ with $x=0.17$ obtained at 23.5 keV.}
\label{f5}
\end{center}
\end{figure}

Figures 5(a)-5(d) show the ZF--$\mu$SR time spectra obtained at 23.5 keV for La$_{2-x}$Ce$_x$CuO$_4$ with $x=0.13$ and 0.17 under various reduction conditions.
For the optimally reduced sample with $x=0.13$, it is found that the relaxation of muon spins gradually becomes fast with decreasing temperature, and fast relaxation is observed at low temperatures, suggesting the development of the Cu-spin correlation.
For $x=0.17$, a fast relaxation is observed for both under- and optimally reduced samples, while the spectra are nearly independent of temperature for the over-reduced sample. 
Here, two significant features are found: (i) for the optimally reduced sample, the overall behavior of the spectra is independent of the Ce concentration $x$, and (ii) for $x=0.17$, the Cu-spin correlation most strongly develops in the optimally reduced sample and is almost paramagnetic down to the base temperature in the over-reduced sample. 

To obtain further information on the Cu-spin correlation at low temperatures, we performed longitudinal-field (LF) $\mu$SR at 2.6 K for optimally reduced LCCO with $x=0.17$. 
The results are shown in Fig. 5(e).
With increasing LF, the asymmetry in a long-time region shows an upward parallel shift, suggesting a static nature of magnetism.
Moreover, at LF $= 100$ G, it appears to exhibit a very slow relaxation in a long-time region. 
Therefore, fluctuating and static spins coexist in the optimally reduced sample with $x=0.17$.
 
The ZF--$\mu$SR time spectra were analyzed using the following equation:

\begin{equation}
A(t) = A_0 {\rm exp}(-\lambda_0 t) {\rm exp}(\frac{-\sigma^2 t^2}{2}) + A_1 {\rm exp}(-\lambda_1 t) + A_{\rm BG}.
\end{equation}
The first and second terms represent slow and fast relaxation components in a region where Cu spins fluctuate fast and the Cu-spin correlation develops, respectively. 
The third is a temperature-independent background term. 
$A_0$, $A_1$, and $A_{\rm BG}$ are initial asymmetries of each component. 
$\lambda_0$ and $\lambda_1$ are the relaxation rates of each exponential function. 
$\sigma$ is the relaxation rate of the Gaussian function in the slow component due to nuclear-dipole fields. 

\begin{figure}
\begin{center}
\includegraphics[width=1.0\linewidth]{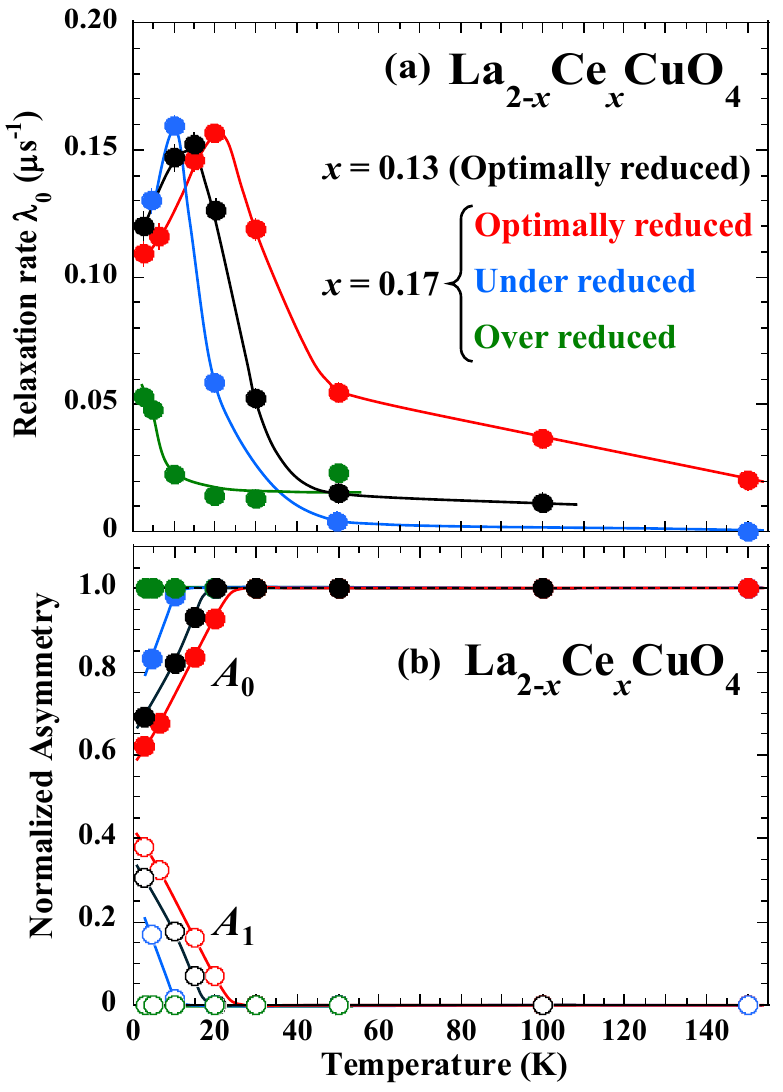}
\caption{(Color online) Temperature dependence of (a) the relaxation rate of muon spins $\lambda_0$ and (b) the initial asymmetries of the slow and fast relaxation components $A_0$ and $A_1$ of La$_{2-x}$Ce$_x$CuO$_4$ with $x=0.13$ and 0.17 under various reduction conditions. Solid lines are guides for the readers' eye.}
\label{f6}
\end{center}
\end{figure}

The temperature dependence of the relaxation rate of the slow component $\lambda_0$ in La$_{2-x}$Ce$_x$CuO$_4$ with $x=0.13$ and 0.17 is shown in Fig. 6(a). 
For samples other than the over-reduced one with $x=0.17$, $\lambda_0$ increases with decreasing temperature, followed by the local maximum, suggesting the development of the Cu-spin correlation at low temperatures. 
The temperature where $\lambda_0$ exhibits the local maximum corresponds to the magnetic transition temperature $T_{\rm m}$ in $\mu$SR. 
In the optimally reduced sample, $T_{\rm m}$ is slightly higher in $x=0.17$ than in $x=0.13$.
For $x=0.17$, $T_{\rm m}$ is slightly higher in the optimally reduced sample than in the under-reduced sample, and the development of the spin correlation is weak in the over-reduced sample.

Figure 6(b) shows the temperature dependence of the initial asymmetries of the slow and fast relaxation components $A_0$ and $A_1$ in La$_{2-x}$Ce$_x$CuO$_4$ with $x=0.13$ and 0.17.
As the spin correlation develops, a fast relaxation component appears in the short-time region of the spectrum, as shown in the optimally reduced $x=0.13$ and 0.17 samples in Fig. 5(a) and 5(c), respectively. 
Consequently, $A_0$ representing the slow relaxation in Eq. (1) decreases, and $A_1$ representing the fast relaxation increases. 
Except for the over-reduced $x=0.17$ sample, $A_0$ decreases at low temperatures. 
The temperature where $A_{\rm 0}$ starts to decrease is close to the peak temperature of $\lambda_0$. 
As $A_0$ decreases, $A_1$ increases. 
As previously mentioned, the sample contains regions with c-axis and (110) orientations, so the magnetic volume fraction cannot be calculated from $A_0$. 

\section{Discussion}

The ZF--$\mu$SR of LCCO thin films revealed the fast relaxation of muon spins corresponding to the development of the Cu-spin correlation at low temperatures for both optimally reduced LCCO with $x=0.13$ and 0.17.
This is contrary to the previous results~\cite{saadaoui} where the fast relaxation of muon spins was observed at $x \le 0.10$. 
On the other hand, the angle-dependent magnetoresistance of LCCO suggested the AF order up to $x=0.14$.~\cite{jin} 
Since angle-dependent magnetoresistance detects the development of static magnetism, AF fluctuations would be observed above $x=0.14$. 
In fact, the present results revealed the presence of the developed Cu-spin correlation at $x=0.13$ and 0.17 in the overdoped regime. 
In PLCCO~\cite{fujita-plcco} and Nd$_{2-x}$Ce$_x$CuO$_4$~\cite{motoyama}, AF fluctuations have been observed in the overdoped regime.
Moreover, our ZF-$\mu$SR of overdoped PLCCO suggested the development of the Cu-spin correlation, but predominant effects of Pr$^{3+}$ moments on the $\mu$SR spectra had to be taken into account.~\cite{malik}
A rough comparison of the relaxation rate of muon spins between PLCCO and LCCO reveals a lower relaxation rate in LCCO than in PLCCO, suggesting that Pr$^{3+}$ moments affect the relaxation rate in PLCCO.

The $\mu$SR spectra for optimally reduced samples with $x=0.13$ and 0.17 show similar temperature dependences. 
In contrast, the behavior of $R_{\rm H}$ is different, which might suggest a decoupling between the electronic system and magnetism. 
However, the behavior of $R_{\rm H}$ is understood in terms of the Fermi-surface topology involving AF fluctuations. 
For $x=0.13$, it has been proposed that because of AF fluctuations, the Fermi surface is reconstructed so as to create electron and hole pockets. 
The sign change of $R_{\rm H}$ with temperature is a result of the competition between these pockets. 
On the other hand, for $x=0.17$, Fermi-surface reconstruction due to AF fluctuations does not occur, and $R_{\rm H}$ is positive over the entire temperature range owing to a large hole Fermi surface. 
Therefore, it is highly probable that the electronic system and magnetism are deeply intertwined.
We suspect that the fundamental magnetic nature is different between the samples with $x=0.13$ and 0.17. 
Specifically, we consider the $\mu$SR spectra for $x=0.13$ to show relaxation due to AF fluctuations, whereas the effects of ferromagnetic fluctuations appear for $x=0.17$, as discussed later. 
However, since $\mu$SR is unable to distinguish between antiferromagnetism and ferromagnetism, this remains speculative.

Although the effects of reduction annealing in the electron-doped cuprates remain unclear, it has been proposed that reduction annealing removes excess oxygen in as-grown samples~\cite{radaelli,schultz} and/or induces oxygen vacancies in the CuO$_2$ plane,~\cite{richard} and that superconductivity emerges from the balance between these effects.~\cite{miyamoto} 
In this study, from depth-resolved $\mu$SR measurements on LCCO thin films under various reduction conditions with $x=0.17$, shown in Fig. 4, information on the magnetic states and the deduced reduction states inside the films was obtained.
The key point is that the development of the Cu-spin correlation is suppressed by oxygen removal. 
The removal of excess oxygen weakens the development of the Cu-spin correlation at low temperatures.~\cite{adachi-musr,adachi-condmat,sumura} 
In addition, oxygen vacancies in the CuO$_2$ plane also hinder the exchange interaction between Cu spins.
For the implantation energy of muons above 13.5 keV, $\mu$SR spectra are independent of the implantation energy, suggesting a homogeneous magnetic state inside the film beyond 60 nm from the surface.
For the under-reduced sample, relaxation is slow at 3.5 keV, indicating that the development of the Cu-spin correlation is weak near the surface of the film.
This is because excess oxygen is removed near the surface more efficiently than deep inside the film.
On the other hand, for the over-reduced sample, relaxation is faster at 3.5 keV than at 13.5 keV, indicating the development of the Cu-spin correlation near the surface.
This suggests the re-inclusion of oxygen near the surface owing to the excessively reduced state in the over-reduced film.
For the optimally reduced sample, the $\mu$SR spectrum near the surface is almost identical to that deep inside the film.

Deep inside the film, the temperature-dependent $\mu$SR spectra of $x=0.17$, shown in Fig. 5, suggest that the Cu-spin correlation developed more in the optimally reduced sample than in the under-reduced sample. 
This is contrary to the effects of reduction annealing on antiferromagnetism, that is, the development of the Cu-spin correlation weakens upon the reduction annealing.~\cite{adachi-musr,adachi-condmat,sumura} 
Therefore, the magnetic state in the optimally reduced $x=0.17$ sample may be different from the AF one. 
For the over-reduced $x=0.17$ sample, further reduction brings about the removal of oxygen in the CuO$_2$ plane, resulting in the destruction of exchange coupling between Cu spins, and the system becomes almost paramagnetic.

Finally, we discuss the development of the Cu-spin correlation in the optimally reduced $x=0.17$ sample. 
The neutron-scattering results of PLCCO indicate that the integrated intensity of $\chi^"$($\omega$) corresponding to low-energy AF fluctuations decreases gradually with overdoping but seems to be finite even at $x \sim 0.20$ where superconductivity disappears. 
This suggests that AF fluctuations are robust even in the nonsuperconducting heavily overdoped regime. 
Therefore, the development of the Cu-spin correlation observed at $x=0.17$ in the present results might be attributed to these AF fluctuations. 
If this is the case, it is necessary to clarify why the spin correlation developed more at $x=0.17$ than at $x=0.13$.

As shown in Fig. 3, the $R_{\rm H}$ of the optimally reduced $x=0.17$ sample is positive in the whole temperature range, suggesting a hole-like Fermi surface without the Fermi-surface reconstruction due to the AF order/fluctuations. 
As mentioned above, the Cu-spin correlation developed more in the optimally reduced sample than in the under-reduced one, which is difficult to understand in terms of the AF-related magnetism.
On the basis of the magnetization curve and magnetoresistance of nonsuperconducting heavily overdoped LCCO with $x \sim 0.18$, it was previously proposed that the observed hysteresis behaviors are a result of the formation of a ferromagnetic order,~\cite{sarkar} which was also proposed in the heavily overdoped regime of hole-doped cuprates.~\cite{kopp,maier,sonier,kurashima,komiyama,adachi-materials} 
Accordingly, the development of the Cu-spin correlation in the optimally reduced $x=0.17$ sample may be related to ferromagnetic order/fluctuations.

\section{Summary}

Using PLD with a Nd:YAG laser, we fabricated overdoped $x=0.13$ and 0.17 LCCO thin films on the STO substrate. 
Although the electrical resistivities of the films are high owing to the incomplete c-axis orientation of the film, $R_{\rm H}$ and $T_{\rm c}$ are in good agreement with former results.~\cite{sawa}
From ZF-- and LF--$\mu$SR measurements using low-energy muons, it was clarified that at $x=0.17$, the under- and over-reduced samples exhibited oxygen removal from the surface of the film and the re-inclusion of oxygen near the surface of the film, respectively. 
It was also found that the Cu-spin correlation developed at low temperatures for both the optimally reduced $x=0.13$ and 0.17 samples.
Although the development of the Cu-spin correlation in the optimally reduced $x=0.17$ sample may be due to AF fluctuations, it may be related to the ferromagnetism that has recently been discussed in high-$T_{\rm c}$ cuprates.
In the future, we plan to investigate the possibility of ferromagnetism through measurements on films with further Ce substitution and through measurements of transport properties.

\begin{acknowledgment}


Helpful advice on the preparation of LCCO thin films by A. Maeda, F. Nabeshima, and I. Tsukada is gratefully acknowledged. 
We thank T. Naoe for the use of the laser microscope at the Japan Atomic Energy Agency to estimate the film thickness.
Some of the resistivity and Hall measurements were performed using PPMS at the CROSS user laboratory. 
The $\mu$SR experiments were performed at the Swiss Muon Source S$\mu$S, Paul Scherrer Institute, Villigen, Switzerland.
This work was supported by JSPS KAKENHI Grants (Nos. JP23108004, JP19H01841 and JP20H05165).

\end{acknowledgment}

\end{document}